\documentclass[twocolumn,tighten]{aastex61}

\newcommand\aastex{AAS\TeX}

\received{}
\revised{}
\accepted{}
\submitjournal{ApJL}

\shorttitle{\aastex\ wavelength-dependent polarization}
\shortauthors{Kataoka et al.}

\begin{document}

\title{The evidence of radio polarization induced by the radiative grain alignment and self-scattering of dust grains in a protoplanetary disk}

\correspondingauthor{Akimasa Kataoka}
\email{akimasa.kataoka@nao.ac.jp}

\author{Akimasa Kataoka}
\altaffiliation{NAOJ fellow}
\altaffiliation{former Humboldt Research Fellow}
\affiliation{National Astronomical Observatory of Japan, Mitaka, Tokyo 181-8588, Japan}
\affiliation{Zentrum f\"ur Astronomie der Universit\"at Heidelberg, Institut f\"ur Theoretische Astrophysik, Albert-Ueberle-Str. 2, D-69120 Heidelberg, Germany}

\author{Takashi Tsukagoshi}
\affiliation{College of Science, Ibaraki University, 2-1-1 Bunkyo, Mito, Ibaraki 310-8512}


\author{Adriana Pohl}
\affiliation{Max Planck Institute for Astronomy, K\"onigstuhl 17, D-69117 Heidelberg, Germany}
\affiliation{Zentrum f\"ur Astronomie der Universit\"at Heidelberg, Institut f\"ur Theoretische Astrophysik, Albert-Ueberle-Str. 2, D-69120 Heidelberg, Germany}

\author{Takayuki Muto}
\affiliation{Division of Liberal Arts, Kogakuin University, 1-24-2 Nishi-Shinjuku, Shinjuku-ku, Tokyo 163-8677}

\author{Hiroshi Nagai}
\affiliation{National Astronomical Observatory of Japan, Mitaka, Tokyo 181-8588, Japan}

\author{Ian W. Stephens}
\affiliation{Harvard-Smithsonian Center for Astrophysics, 60 Garden Street, Cambridge, MA 02138, USA}

\author{Kohji Tomisaka}
\affiliation{National Astronomical Observatory of Japan, Mitaka, Tokyo 181-8588, Japan}

\author{Munetake Momose}
\affiliation{College of Science, Ibaraki University, 2-1-1 Bunkyo, Mito, Ibaraki 310-8512}



\begin{abstract}
The mechanisms causing millimeter-wave polarization in protoplanetary disks are under debate.
To disentangle the polarization mechanisms, we observe the protoplanetary disk around HL Tau at 3.1 mm with the Atacama Large Millimeter/submillimeter Array (ALMA), which had polarization detected with CARMA at 1.3 mm.
We successfully detect the ring-like azimuthal polarized emission at 3.1 mm.
This indicates that dust grains are aligned with the major axis being in the azimuthal direction, which is consistent with the theory of radiative alignment of elongated dust grains, where the major axis of dust grains is perpendicular to the radiation flux.
Furthermore, the morphology of the polarization vectors at 3.1 mm is completely different from those at 1.3 mm.
We interpret that the polarization at 3.1 mm to be dominated by the grain alignment with the radiative flux producing azimuthal polarization vectors, while the self-scattering dominates at 1.3 mm and produces the polarization vectors parallel to the minor axis of the disk. 
By modeling the total polarization fraction with a single grain population model, the maximum grain size is constrained to be $100{\rm~\mu m}$, which is smaller than the previous predictions based on the spectral index between ALMA at 3 mm and VLA at 7 mm.
\end{abstract}




\section{Introduction} \label{sec:intro}

The millimeter-wave polarization of circumstellar disks is a powerful tool to investigate the grain properties.
However, the mechanisms that produce  the millimeter-wave polarization are under debate.

There have been two mechanisms proposed that can explain millimeter-wave polarization in protoplanetary disks: the alignment and scattering of dust grains.
The polarized thermal dust emission has been used as a tracer of magnetic fields in scales of molecular clouds and star-forming regions \citep[e.g.,][]{Lai01, Lai02, Girart06, Girart09, Rao09, Stephens13, Hull13, Hull14, Cortes16, PlanckXXXIII,Ward-Thompson17} or circumstellar disks around Class 0-I protostars \citep{Rao14, Segura-Cox15}.
It is considered that the major axis of elongated dust grains is aligned with the direction perpendicular to magnetic fields with a help of radiative torque \citep[e.g.,][]{LazarianHoang07}.
In protoplanetary disks, if the dust grains are aligned with magnetic fields, the polarization fraction is expected to be high enough to detect with interferometers \citep{ChoLazarian07, Bertrang17}.
However, there were many attempts to detect polarized emission from protoplanetary disks around Class II or III protostars, which resulted in non-detections \citep{Hughes09, Hughes13}.
It has recently been pointed out that dust grains in disks may not align with magnetic fields but with radiation fields \citep{Tazaki17}.

The other possibility of the polarization mechanisms is the self-scattering of the thermal dust emission \citep{Kataoka15}.
If the grain size is comparable to the wavelengths, scattering-induced polarization can produce 2-3 \% polarization from protoplanetary disks.
If it is the case, we can constrain the grain size from the polarization fraction.
Together with dust coagulation theory, we can test the grain growth theory with polarization observations \citep{Pohl16}.

There have been two resolved detections of mm-wave polarization for disks that are Class I or older.
The first detection was made on the protoplanetary disk around HL Tau with CARMA at 1.3 mm and SMA at 0.87 mm \citep{Stephens14}.
The polarization vectors are parallel to the direction of the minor axis of the disk. 
This morphology has been first interpreted as complex magnetic fields dominated by toroidal components \citep[see also][]{Matsakos16}.
However, the polarimetric image can also be interpreted with the self-scattering \citep{Kataoka16a, Yang16a}.
The other polarization detection is from the disk around HD 142527, which has conspicuous asymmetric ring emission of dust continuum at the wavelength of 0.87 mm \citep{Kataoka16b}.
The polarization vectors are mainly directed radially, but are directed azimuthally in the outer region of the disk. 
The flip of the polarization vectors are expected from the self-scattering theory \citep{Kataoka15}.
This confirms that the self-scattering is working on the protoplanetary disk.
However, it does not exclude the possibility of the grain alignment.

To disentangle the mechanisms between the grain alignment and self-scattering, multi-wave polarization observations are essential.
The wavelength dependence of the polarization fraction is not strong in the case of the grain alignment while it is strong in the case of the self-scattering because the scattering-induced polarization is efficient only when the maximum grain size is around $\lambda/2\pi$ where $\lambda$ is the wavelengths \citep{Kataoka15}.

To obtain the wavelength-dependent polarimetric images, we observe the HL Tau disk with Atacama Large Millimeter/submillimeter Array (ALMA) with using Band 3.
HL Tau is a young star in Taurus molecular cloud with the distance of 140 pc \citep{Rebull04}.
The circumstellar disk is around in $\sim 100$ AU scale \citep{Kwon11}.
The disk has several ring and gap structures with tens of AU scales \citep{Partnership15}.
The observed band corresponds to wavelengths of 3.1 mm, which is sufficiently longer than the previous CARMA polarimetric observations at 1.3 mm \citep{Stephens14}.

\section{Observations} \label{sec:obs}

HL Tau was observed by ALMA on October 12, 2016 during its Cycle 4 operation (2016.1.00115.S, PI: A.Kataoka). 
The antenna configuration was C40-6, and 41 antennas were operating.
The correlater processed four spectral windows centered at $90.5, 92.5, 102.5,$ and $104.5$ GHz with a bandwidth of 1.75 GHz each.
The bandpass, amplitude and phase were calibrated by observations of J0510+1800, J0423-0120, and J0431+1731, respectively, and the polarization calibration was performed by observations of J0510+1800.
The raw data were reduced by the EA-ARC staff.

We further perform the iterative CLEAN deconvolution imaging with self-calibration to improve the image quality.
We employ the briggs weighting with the robust parameter of $0.5$ and the multiscale option with scale parameters of 0, 0.3, 0.9 arcsec.
The beam size of the final product is $0.45\arcsec \times 0.29\arcsec$, corresponding to $\sim 63 \times 41$ AU at the distance of 140 pc to the target.
The rms for Stokes I, Q, U is 9.6, 6.9, 6.9 $\mu$Jy, respectively.

\section{Results} \label{sec:results}
\begin{figure*}[ht!]
\epsscale{1.8}
\plottwo{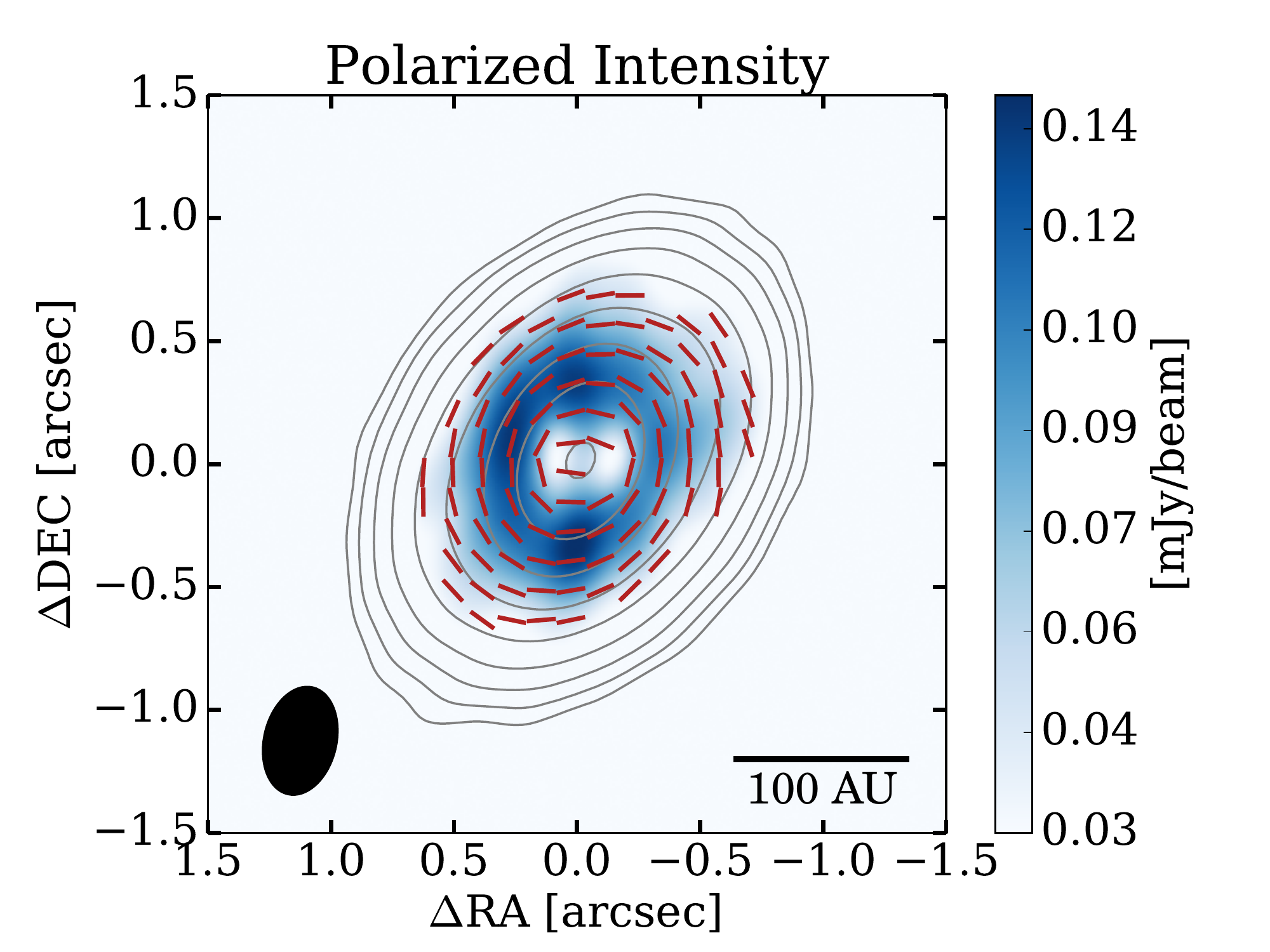}{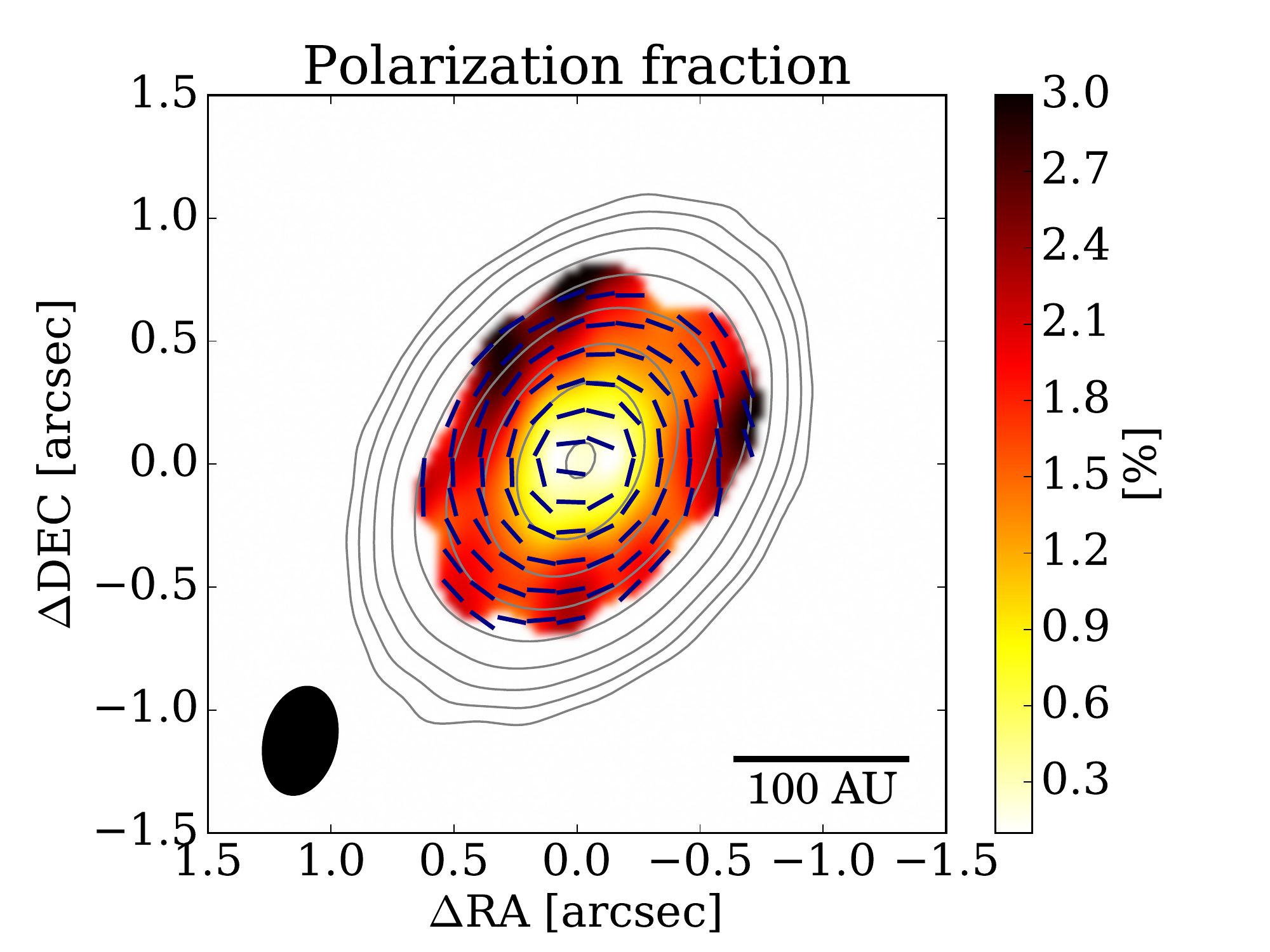}
\caption{
ALMA Band 3 observations of the HL Tau disk. 
The wavelength is 3.1 mm.
The top panel shows the polarized intensity in color scale, the polarization direction in red vectors, and the continuum intensity in the solid contour.
The vectors are shown where the polarized intensity is larger than $5 \sigma_{\rm PI}$. 
The contours corresponds to $(10,20,40,80,160,320,640,1280) \times$ the rms of 9.6 $\mu$Jy.
The bottom panel shows that the polarization fraction in color scale, polarization vectors in red, and the same continuum intensity contours as the top.
}
\label{fig:obs}
\end{figure*}

The top panel of Fig. \ref{fig:obs} shows the polarized intensity in color scale overlaid with polarization vectors \footnote{We plot the polarization vectors not scaling with the polarization fraction but written with the same length because this allows for the polarization morphology to be more obvious.
However, the reliability depends not on the polarization fraction but on the polarized intensity.}, and the contour represents the continuum emission.
The bottom panel of Fig. \ref{fig:obs} shows the polarization fraction in color scale and the others are the same as the top panel.
Due to the lower spatial resolution than the long baseline campaign \citep{Partnership15}, the multiple ring and gap structure of the continuum is not resolved.
The total flux density is 75.1 mJy, which is consistent with the previous ALMA observations with Band 3 \citep[74.3 mJy;][]{Partnership15}.

We successfully detect the ring-like polarized emission at 3.1 mm.
The polarized intensity has a peak of 145 $\mu$Jy/beam, which corresponds to 21 sigma detection with the rms of 6.9 $\mu$Jy.
The peak of the polarized intensity is not located at the central star but on the ring.
We see three blobs on the ring but it may be due to the interferometric effects.
The polarized intensity at the location of the central star is lower than the other regions.
We interpret this structure as beam dilution of the central region where polarization is expected to be azimuthal and thus canceled out to each other.
The polarization fraction is around 1.8 \% on the ring.

The flux densities of the entire disk are $-39.7$ $\mu$Jy for Stokes Q and $-40.6$ $\mu$Jy for Stokes U. 
Therefore, the integrated polarized intensity is $PI=\sqrt{Q^2 + U^2 - \sigma_{\rm PI}^2}=56.4$ $\mu$Jy.
Dividing the total polarized intensity by the total Stokes I, we obtain 0.08 \% for the total polarization fraction.
The instrumental polarization contamination of the ALMA interferometers is the polarization fraction of 0.1 \% for a point source in the center of the field or 0.3 \% within up to the inner 1/3 of the FWHM (see the technical handbook of ALMA. More discussion is found in \citealt{Nagai16}).
The derived polarization fraction of the integrated flux corresponds to the case of the point source.
Therefore, the upper limit of the integrated polarization fraction of the HL Tau disk at 3.1 mm by our observations is 0.1 \%.
 The low total polarization fraction means that we could not have detected polarization if we had not resolved the target.

\section{Discussions}

\subsection{Comparison with the previous observations}
\label{sec:mechanism}

\begin{figure*}[ht!]
\epsscale{1.1}
\plottwo{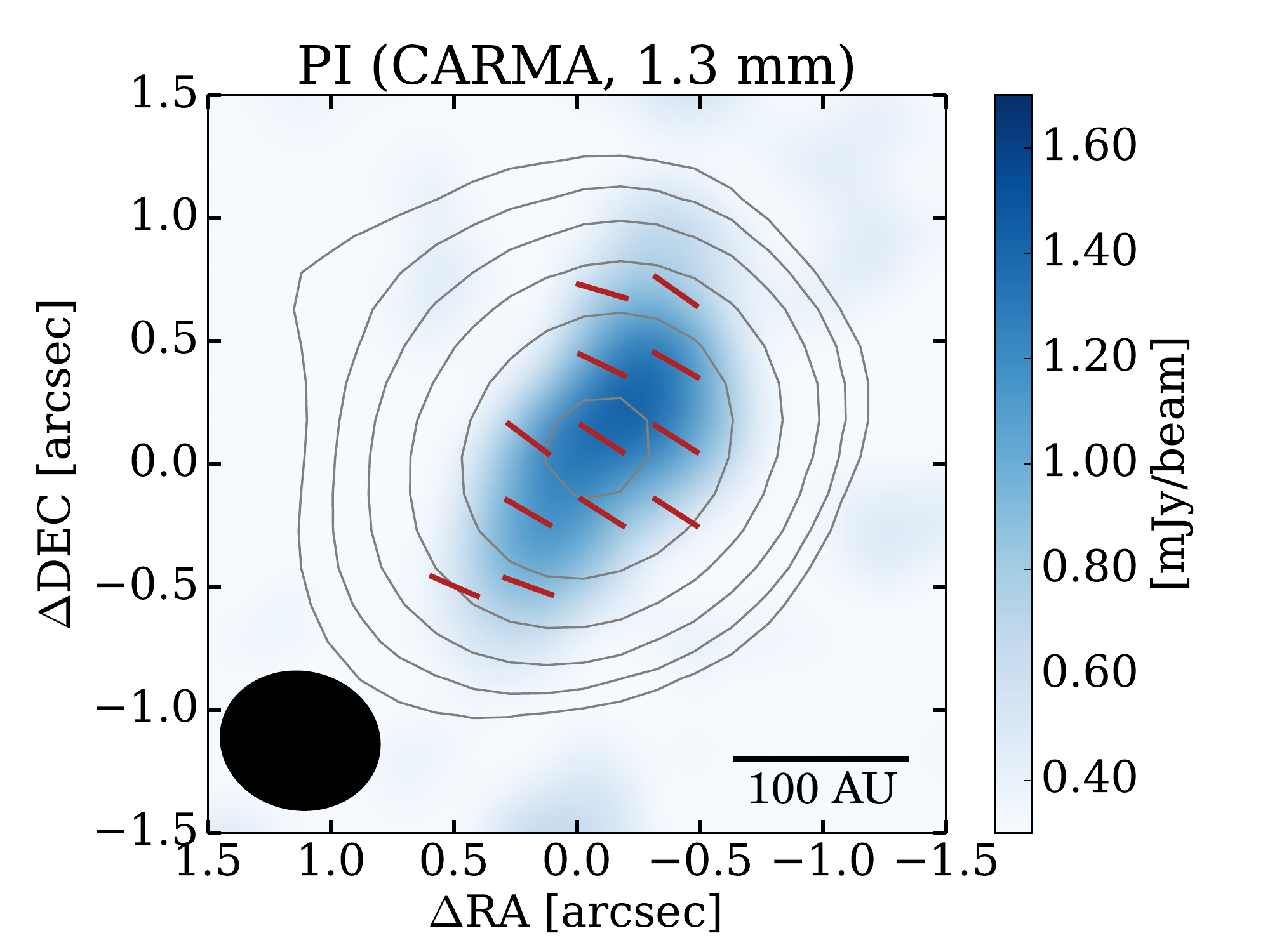}{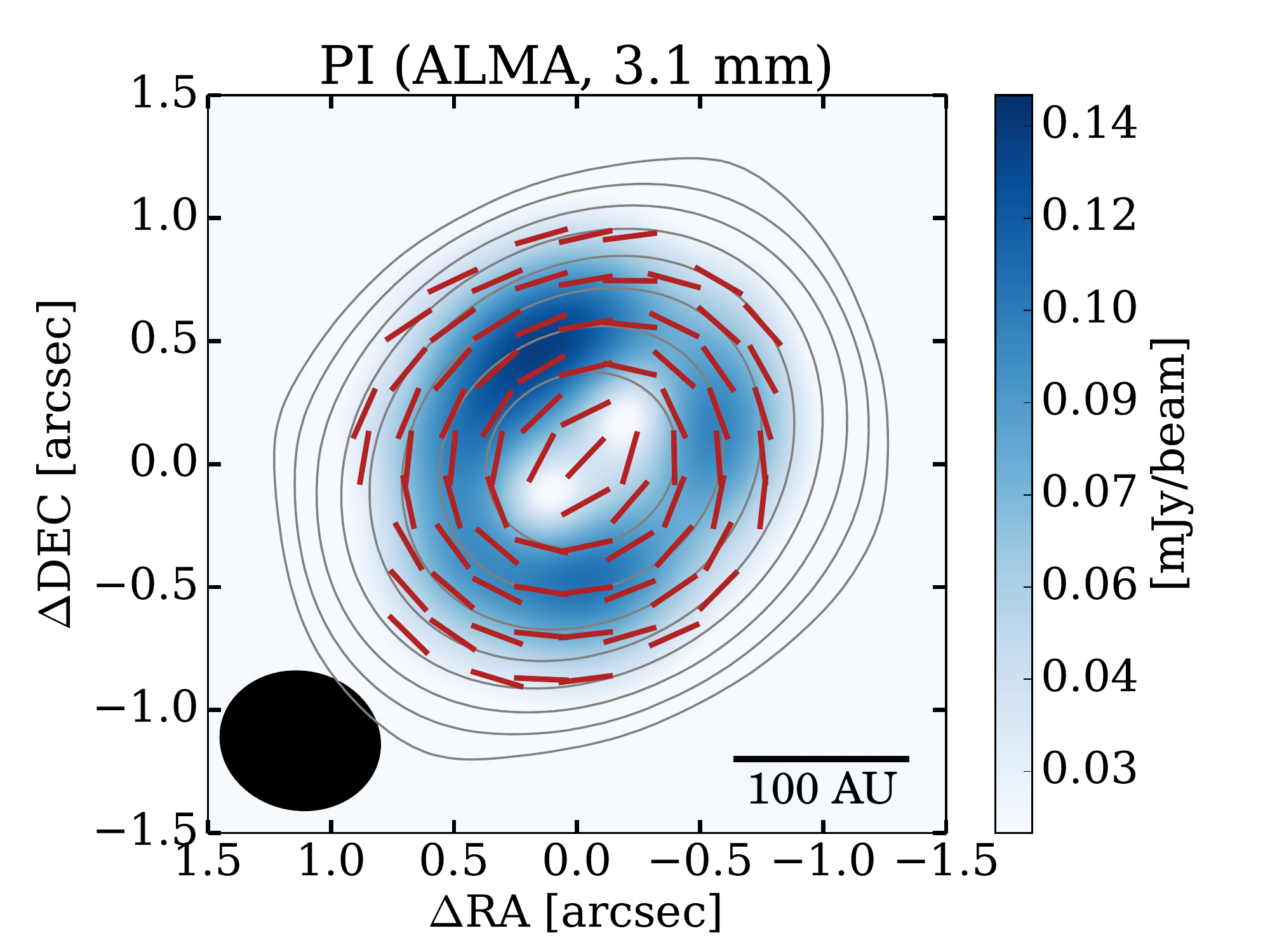}
\caption{Comparison of the polarization images between $\lambda=1.3$ mm \citep[CARMA][]{Stephens14} and $\lambda=3.1$ mm (ALMA, this observation).
The ALMA image is smoothed to have the same beam size of CARMA, where the beam size is $0.65\arcsec \times 0.56\arcsec$ with the PA of 79.5 degrees.
The color scale represents the polarized intensity while the grey contours represent the continuum emission.
The levels of the grey contours are $(10,20,40,80,160,320,640,1280) \times \sigma_{\rm I}$ where $\sigma_{\rm I}=2.1$ mJy/beam for the CARMA data and  $\sigma_{\rm I}=0.017$ mJy/beam ALMA data. 
}
\label{fig:compareCARMA}
\end{figure*}

To interpret the polarization emission from HL Tau, we compare the ALMA results with the previous observations.
We show the CARMA polarization observations by \citealt{Stephens14} in Fig. \ref{fig:compareCARMA}(a), where we show the polarization vectors, while the original paper presents vectors rotated by 90 degrees to show inferred magnetic fields (i.e., if polarization is due to grains aligned with the magnetic field).
For the comparison, we smooth the ALMA observations with the beam size of $0.65\arcsec \times 0.56\arcsec$ and PA of 79.5 degrees, which is the beam size of the CARMA observations \citep{Stephens14}, as shown in Fig. \ref{fig:compareCARMA}(b).

The 3.1-mm polarization morphology with our ALMA observation is completely different from that at 1.3 mm.
At 1.3 mm, the polarization vectors show the direction parallel to the minor axis.
At 3.1 mm, however, the polarization vectors show the circular pattern.

\begin{figure*}[ht!]
\epsscale{1.2}
\plotone{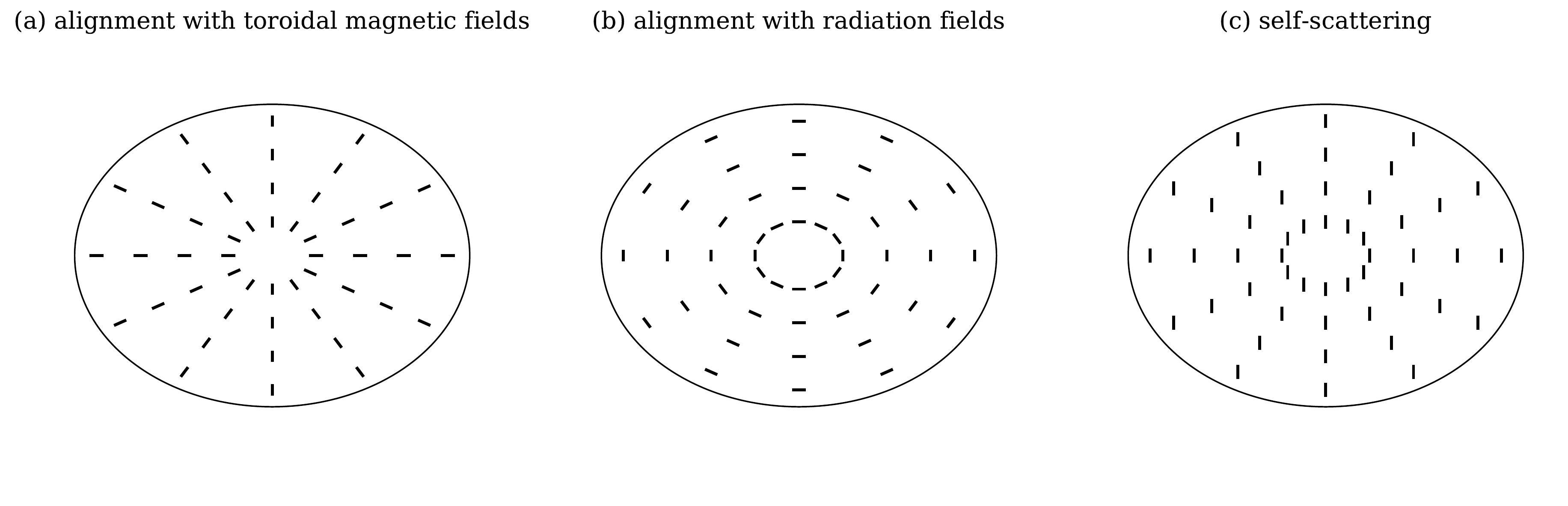}
\caption{
Schematic illustrations for the differences of polarization vectors of each mechanism of polarization of thermal dust emission.
The major axis is in the horizontal direction.
Note that each panel represents E-vectors.
(a) Grain alignment with the toroidal magnetic fields.
(b) Grain alignment with the radiation fields.
(c) Self-scattering of the thermal dust emission}
\label{fig:schematic}
\end{figure*}

To interpret the strong dependence of the polarization on the wavelengths, we consider three possibilities: alignment by magnetic fields, alignment by radiation anisotropy, and self-scattering of thermal dust emission.
Figure \ref{fig:schematic} shows the schematic illustration of the polarization vectors with the three different mechanisms.
Figure \ref{fig:schematic} (a) shows the case of the grain alignment with toroidal magnetic fields \citep[e.g.,][]{ChoLazarian07}.
As we expect the major axis of the grains is aligned perpendicular to the magnetic fields.
In the existence of the toroidal magnetic fields \citep[e.g.,][]{Brandenburg95}, the polarization vectors are in the radial direction.
Figure \ref{fig:schematic} (b) shows the case of the grain alignment with radiation anisotropies \citep[e.g.,][]{Tazaki17}.
The flux gradient is in general in the outgoing radial direction.
Considering that the major axis of the grains is perpendicular to the flux gradient, the polarization vectors should be in the azimuthal direction.
Figure \ref{fig:schematic} (c) shows the case of self-scattering \citep{Kataoka15, Kataoka16a, Yang16a}.
The flux coming parallel to the major axis is much stronger than that parallel to the minor axis.
This leads to the polarization vectors parallel to the minor axis \footnote{While we plot the polarization vectors at the central part of the disk \citep{Kataoka16a}, the polarization vectors might become azimuthal at the outer edge of the disk because the flux gradient is stronger than the quadrupole components \citep{Pohl16, Yang16a}}.

In the case of HL Tau, the 3.1 mm polarimetric image is consistent with case (b), which is the alignment with the radiation anisotropy.
The polarization vectors are essentially perpendicular to that expected for case (a), and thus we can rule out grain alignment with the toroidal magnetic fields at 3.1 mm.
However, we cannot rule out the grain alignment with the poloidal magnetic fields.
In the case of 1.3 mm image, however, we interpret it with (c) the self-scattering of the thermal dust emission, which provides the polarization vectors parallel to the minor axis.
However, there could be also some contributions of (b) the alignment with the radiation fields to the polarization, which enhances the polarization vectors at the north-west and south-east regions (along the major axis) while decreases the polarization fraction at the north-east and south-west regions (along the minor axis).

The wavelength dependence in the polarization fraction in the case of the self-scattering is strong \citep{Kataoka15} while it is weaker in the case of the grain alignment. 
Therefore, the most natural interpretation is that the alignment with the radiation fields provides the axisymmetric azimuthal polarization vectors on both wavelengths while the self-scattering dominates at 1.3 mm.

\subsection{Modeling the scattered components}

By modeling the scattered components of the polarization, we can constrain the grain size in the HL Tau disk.
To model the scattering components in polarization, we consider the total polarization fraction across HL Tau. 
If we integrate the polarization all over the disk, the axisymmetric vectors are canceled out.
The scattering-induced polarization provides the vectors parallel to the minor axis, which resides as the total polarization fraction.
However, the alignment with the radiative flux is almost axisymmetric and thus does not contribute so much on the integrated polarization fraction.
We estimate the contribution of the radiative flux alignment to the total polarization fraction assuming that the disk is geometrically and optically thin, the local alignment efficiency $p$ is the same in the entire disk \citep{FiegePudritz00, Tomisaka11}, and there is no wavelength dependence.
The contribution is calculated to be $0.114\times p$ and the polarization vectors are parallel to the major axis.

We have already discussed that the upper limit of the total polarization fraction is 0.1 \% at 3.1 mm with our ALMA observations.
The polarization fraction with SMA is reported to be $0.86 \pm 0.4\%$ at 0.87 mm \citep{Stephens14}.
Note that the detection was 2 sigma significance, which might be an upper limit of the polarization fraction while we use the reported value in \citealt{Stephens14} in this paper.
We calculate the total degree of polarization observed with CARMA at 1.3 mm with the data reported by \citealt{Stephens14}, which is $0.52 \pm 0.1 \%$.

\begin{figure*}[ht!]
\plotone{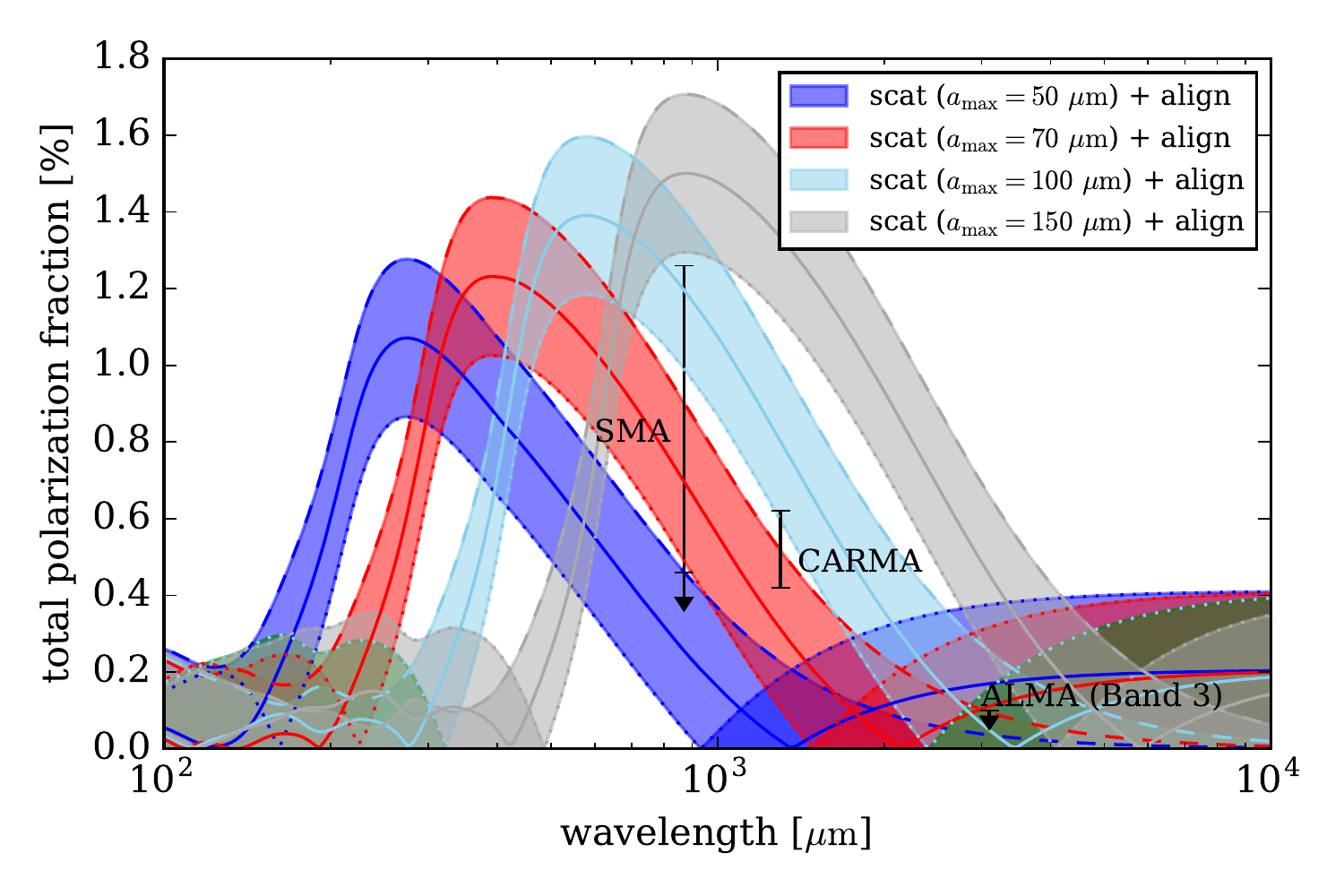}
\caption{The total polarization fraction as a function of the observed wavelengths.
The total polarization fraction is derived by integrating each Stokes I, Q, and U component.
The curves represent the prediction of the HL Tau disk with the self-scattering model where the maximum grain sizes are $a_{\rm max}= $ 50, 70, 100, and $150{\rm~\mu m}$ \citep{Kataoka16a} and with the radiative alignment model.
The dashed, solid, and dotted lines represent the models with the local alignment efficiency of $p=0,1.8, 3.6\%$, respectively.
}
\label{fig:wavedependence}
\end{figure*}

Figure \ref{fig:wavedependence} compares that the theoretical prediction and the observational results of the total polarization fraction.
The contribution of the self-scattering is estimated as $P = CP_{90}\omega$, where C is the calibration factor set to be 2.0 \%, $P_{90}$ is the polarization efficiency for the scattering angle of 90 degrees in a single scattering, and $\omega$ is the albedo \citep{Kataoka15, Kataoka16a}.
This prediction is confirmed to match the results radiative transfer calculations of the polarization due to the self-scattering within the error of 50 \% \citep{Kataoka16a}.
The dust grains are assumed to be spherical and have a power-law size distribution of $n(a)\propto a^{-3.5}$.
We vary the maximum grain size $a_{\rm max}$ for $a_{\rm max}= $ 50, 70, 100 and $150{\rm~\mu m}$.
The contribution of the radiative alignment is estimated in three cases where the local polarization fraction is $p=0,1.8$, and $3.6 \%$.
We choose $p=1.8 \%$ for the fiducial case based on our observations, but the local polarization fraction could be higher if the spatial resolution is better or smaller if it is diluted by the beam.
Then, we obtain the contribution to the total polarization fraction of $0.114\times p=0, 0.21,$ and $0.41$, respectively.

To explain the results of SMA and CARMA, the self-scattering is essential. 
In the case of  $a_{\rm max}=50 {\rm~\mu m}$, the expected polarization fraction is too low to explain the observations because the albedo is too small at the wavelengths of 0.87 and 1.3 mm. 
In the case of  $a_{\rm max}=150 {\rm~\mu m}$, on the other hand, the polarization fraction is too high to explain the SMA, CARMA, and ALMA observations.
In the case of  $a_{\rm max}=70 {\rm~\mu m}$, if the contribution from the radiative alignment is small in the range of $0<p<1.8\%$, the observations can be explained. 
However, if the local alignment efficiency produces $p>1.8\%$, the total polarization fraction is more than 0.1 \% at 3.1 mm, which is not consistent with the upper limit of the ALMA observations. 
In the case of  $a_{\rm max}=100 {\rm~\mu m}$, the combination of the self-scattering and the radiative alignment greatly explains the whole observations.
Therefore, we conclude that the maximum grain size is constrained to be $a_{\rm max}=100 {\rm~\mu m}$ from the polarimetric observations of SMA, CARMA, and ALMA.

\subsection{Dust grains in the HL Tau disk}

The opacity index of dust grains with $a_{\rm max}=100 {\rm~\mu m}$ is the same as the interstellar medium or higher, which corresponds to $\beta \sim 1.7$ or more for the standard mixture of silicate, water ice, or carbonaceous materials \citep[e.g.,][]{MiyakeNakagawa93, Ricci10a}. 
However, the HL Tau disk has been observed at 7 mm using VLA, and spectral index between 3 mm with ALMA and 7 mm with VLA is around $3.0$ \citep{Carrasco-Gonzalez16}.
The continuum emissions at 3 mm and 7 mm are optically thin, and thus the opacity index is $\beta \sim 1.0$, which is not consistent with the value inferred from our polarization observations.

One possibility that could resolve this problem would be to consider two dust population. 
One dust polarization consists primarily of small grains (10s of $\mu$m in size) that dominate the polarization in the disk. 
The second dust population primarily consists of large grains ($\gg$100 $\mu$m in size) that produce significant continuum emission but negligible scattering at these wavelengths.

Introducing porosity does not solve the problem.
The absorption opacity at millimeter wavelengths is determined by mass-to-area ratio \citep{Kataoka14}. 
To explain the small opacity index, we need more massive aggregates if we consider porous dust aggregates.
On the other hand, the scattering properties at long wavelengths are determined by the size of the whole aggregates \citep{Tazaki16, Min16}.
Therefore, the dust aggregates larger than $a_{\rm max}=100 {\rm~\mu m}$ produces much more degree of polarization emission at 3.1 mm, which is not consistent with our observations.

Another possibility is to change the constituent materials that determine the refractive index at the millimeter wavelengths.
The grain size constraints from the polarization are mainly determined by the wavelength dependence of the albedo, which is determined by the combination of the grain size and observed wavelengths, which does not so much change even if the refractive index is changed. 
On the other hand, the opacity slope of the dust grains is proportional to the imaginary part of the refractive index at the observed wavelengths.
Therefore, it is possible to reconcile the problem by considering a different mixture of grains than those commonly assumed \citep[e.g.,][]{Woitke16}.

There are still plenty of parameters that could affect the total polarization fraction: wavelength dependence of the alignment efficiency, optical depth effects, or vertical structure of the disk, etc.
Further detailed modeling should be done in future papers.

The small grain sizes that are required to produce the polarization by scattering has huge implications on the mechanisms to form the ring and gap structures of the HL Tau disk \citep{Partnership15}.
The inferred grain size suggests that the grains are more coupled with the gas than grains with size of 1 mm or larger.
For example, the disk turbulence should be extremely weak because the dust settling is required to reproduce the geometrically thin dust disk \citep[e.g.,][]{Pinte16}.
Furthermore, there are some scenarios that requires the dust to be decoupled from the gas such as trapping dust grains at gas pressure bumps produced by planets \citep{Dipierro15} and the secular gravitational instability of the gas and dust disk \citep{TakahashiInutsuka14}.
In the case of the sintering-induced fragmentation at around snowlines, the grain size is determined by the fragmentation properties \citep{Okuzumi16}, which is not tested with various parameters.
The mm-wave polarization requests these scenarios to produce $100 {\rm~\mu m}$ grains.

\acknowledgments
We appreciate the discussions with Cornelis P. Dullemond, Tomoyuki Hanawa, Satoshi Okuzumi, and Benjamin Wu.
This work is supported by JSPS KAKENHI Grant Numbers JP15K17606, JP26800106, JP17H01103, JP15H02074.
T.M. is supported by NAOJ ALMA Scientific Research Grant Numbers 2016-02A.
This paper makes use of the following ALMA data: ADS/JAO.ALMA\#2016.1.00115.S. ALMA is a partnership of ESO (representing its member states), NSF (USA) and NINS (Japan), together with NRC (Canada) and NSC and ASIAA (Taiwan) and KASI (Republic of Korea), in cooperation with the Republic of Chile. The Joint ALMA Observatory is operated by ESO, AUI/NRAO and NAOJ.

\newcommand{\SortNoop}[1]{}


\begin{thebibliography}{}
\expandafter\ifx\csname natexlab\endcsname\relax\def\natexlab#1{#1}\fi

\bibitem[{{ALMA Partnership} {et~al.}(2015){ALMA Partnership}, {Brogan},
  {P{\'e}rez}, {Hunter}, {Dent}, {Hales}, {Hills}, {Corder}, {Fomalont},
  {Vlahakis}, {Asaki}, {Barkats}, {Hirota}, {Hodge}, {Impellizzeri}, {Kneissl},
  {Liuzzo}, {Lucas}, {Marcelino}, {Matsushita}, {Nakanishi}, {Phillips},
  {Richards}, {Toledo}, {Aladro}, {Broguiere}, {Cortes}, {Cortes}, {Espada},
  {Galarza}, {Garcia-Appadoo}, {Guzman-Ramirez}, {Humphreys}, {Jung}, {Kameno},
  {Laing}, {Leon}, {Marconi}, {Mignano}, {Nikolic}, {Nyman}, {Radiszcz},
  {Remijan}, {Rod{\'o}n}, {Sawada}, {Takahashi}, {Tilanus}, {Vila Vilaro},
  {Watson}, {Wiklind}, {Akiyama}, {Chapillon}, {de Gregorio-Monsalvo}, {Di
  Francesco}, {Gueth}, {Kawamura}, {Lee}, {Nguyen Luong}, {Mangum}, {Pietu},
  {Sanhueza}, {Saigo}, {Takakuwa}, {Ubach}, {van Kempen}, {Wootten},
  {Castro-Carrizo}, {Francke}, {Gallardo}, {Garcia}, {Gonzalez}, {Hill},
  {Kaminski}, {Kurono}, {Liu}, {Lopez}, {Morales}, {Plarre}, {Schieven},
  {Testi}, {Videla}, {Villard}, {Andreani}, {Hibbard}, \&
  {Tatematsu}}]{Partnership15}
{ALMA Partnership}, {Brogan}, C.~L., {P{\'e}rez}, L.~M., {et~al.} 2015, \apjl,
  808, L3

\bibitem[{{Bertrang} {et~al.}(2017){Bertrang}, {Flock}, \& {Wolf}}]{Bertrang17}
{Bertrang}, G.~H.-M., {Flock}, M., \& {Wolf}, S. 2017, \mnras, 464, L61

\bibitem[{{Brandenburg} {et~al.}(1995){Brandenburg}, {Nordlund}, {Stein}, \&
  {Torkelsson}}]{Brandenburg95}
{Brandenburg}, A., {Nordlund}, A., {Stein}, R.~F., \& {Torkelsson}, U. 1995,
  \apj, 446, 741

\bibitem[{{Carrasco-Gonz{\'a}lez} {et~al.}(2016){Carrasco-Gonz{\'a}lez},
  {Henning}, {Chandler}, {Linz}, {P{\'e}rez}, {Rodr{\'{\i}}guez},
  {Galv{\'a}n-Madrid}, {Anglada}, {Birnstiel}, {van Boekel}, {Flock}, {Klahr},
  {Macias}, {Menten}, {Osorio}, {Testi}, {Torrelles}, \&
  {Zhu}}]{Carrasco-Gonzalez16}
{Carrasco-Gonz{\'a}lez}, C., {Henning}, T., {Chandler}, C.~J., {et~al.} 2016,
  \apjl, 821, L16

\bibitem[{{Cho} \& {Lazarian}(2007)}]{ChoLazarian07}
{Cho}, J., \& {Lazarian}, A. 2007, \apj, 669, 1085

\bibitem[{{Cortes} {et~al.}(2016){Cortes}, {Girart}, {Hull}, {Sridharan},
  {Louvet}, {Plambeck}, {Li}, {Crutcher}, \& {Lai}}]{Cortes16}
{Cortes}, P.~C., {Girart}, J.~M., {Hull}, C.~L.~H., {et~al.} 2016, \apjl, 825,
  L15

\bibitem[{{Dipierro} {et~al.}(2015){Dipierro}, {Price}, {Laibe}, {Hirsh},
  {Cerioli}, \& {Lodato}}]{Dipierro15}
{Dipierro}, G., {Price}, D., {Laibe}, G., {et~al.} 2015, \mnras, 453, L73

\bibitem[{{Fiege} \& {Pudritz}(2000)}]{FiegePudritz00}
{Fiege}, J.~D., \& {Pudritz}, R.~E. 2000, \apj, 544, 830

\bibitem[{{Girart} {et~al.}(2009){Girart}, {Beltr{\'a}n}, {Zhang}, {Rao}, \&
  {Estalella}}]{Girart09}
{Girart}, J.~M., {Beltr{\'a}n}, M.~T., {Zhang}, Q., {Rao}, R., \& {Estalella},
  R. 2009, Science, 324, 1408

\bibitem[{{Girart} {et~al.}(2006){Girart}, {Rao}, \& {Marrone}}]{Girart06}
{Girart}, J.~M., {Rao}, R., \& {Marrone}, D.~P. 2006, Science, 313, 812

\bibitem[{{Hughes} {et~al.}(2013){Hughes}, {Hull}, {Wilner}, \&
  {Plambeck}}]{Hughes13}
{Hughes}, A.~M., {Hull}, C.~L.~H., {Wilner}, D.~J., \& {Plambeck}, R.~L. 2013,
  \aj, 145, 115

\bibitem[{{Hughes} {et~al.}(2009){Hughes}, {Wilner}, {Cho}, {Marrone},
  {Lazarian}, {Andrews}, \& {Rao}}]{Hughes09}
{Hughes}, A.~M., {Wilner}, D.~J., {Cho}, J., {et~al.} 2009, \apj, 704, 1204

\bibitem[{{Hull} {et~al.}(2013){Hull}, {Plambeck}, {Bolatto}, {Bower},
  {Carpenter}, {Crutcher}, {Fiege}, {Franzmann}, {Hakobian}, {Heiles}, {Houde},
  {Hughes}, {Jameson}, {Kwon}, {Lamb}, {Looney}, {Matthews}, {Mundy}, {Pillai},
  {Pound}, {Stephens}, {Tobin}, {Vaillancourt}, {Volgenau}, \&
  {Wright}}]{Hull13}
{Hull}, C.~L.~H., {Plambeck}, R.~L., {Bolatto}, A.~D., {et~al.} 2013, \apj,
  768, 159

\bibitem[{{Hull} {et~al.}(2014){Hull}, {Plambeck}, {Kwon}, {Bower},
  {Carpenter}, {Crutcher}, {Fiege}, {Franzmann}, {Hakobian}, {Heiles}, {Houde},
  {Hughes}, {Lamb}, {Looney}, {Marrone}, {Matthews}, {Pillai}, {Pound},
  {Rahman}, {Sandell}, {Stephens}, {Tobin}, {Vaillancourt}, {Volgenau}, \&
  {Wright}}]{Hull14}
{Hull}, C.~L.~H., {Plambeck}, R.~L., {Kwon}, W., {et~al.} 2014, \apjs, 213, 13

\bibitem[{{Kataoka} {et~al.}(2016{\natexlab{a}}){Kataoka}, {Muto}, {Momose},
  {Tsukagoshi}, \& {Dullemond}}]{Kataoka16a}
{Kataoka}, A., {Muto}, T., {Momose}, M., {Tsukagoshi}, T., \& {Dullemond},
  C.~P. 2016{\natexlab{a}}, \apj, 820, 54

\bibitem[{{Kataoka} {et~al.}(2014){Kataoka}, {Okuzumi}, {Tanaka}, \&
  {Nomura}}]{Kataoka14}
{Kataoka}, A., {Okuzumi}, S., {Tanaka}, H., \& {Nomura}, H. 2014, \aap, 568,
  A42

\bibitem[{{Kataoka} {et~al.}(2015){Kataoka}, {Muto}, {Momose}, {Tsukagoshi},
  {Fukagawa}, {Shibai}, {Hanawa}, {Murakawa}, \& {Dullemond}}]{Kataoka15}
{Kataoka}, A., {Muto}, T., {Momose}, M., {et~al.} 2015, \apj, 809, 78

\bibitem[{{Kataoka} {et~al.}(2016{\natexlab{b}}){Kataoka}, {Tsukagoshi},
  {Momose}, {Nagai}, {Muto}, {Dullemond}, {Pohl}, {Fukagawa}, {Shibai},
  {Hanawa}, \& {Murakawa}}]{Kataoka16b}
{Kataoka}, A., {Tsukagoshi}, T., {Momose}, M., {et~al.} 2016{\natexlab{b}},
  \apjl, 831, L12

\bibitem[{{Kwon} {et~al.}(2011){Kwon}, {Looney}, \& {Mundy}}]{Kwon11}
{Kwon}, W., {Looney}, L.~W., \& {Mundy}, L.~G. 2011, \apj, 741, 3

\bibitem[{{Lai} {et~al.}(2001){Lai}, {Crutcher}, {Girart}, \& {Rao}}]{Lai01}
{Lai}, S.-P., {Crutcher}, R.~M., {Girart}, J.~M., \& {Rao}, R. 2001, \apj, 561,
  864

\bibitem[{{Lai} {et~al.}(2002){Lai}, {Crutcher}, {Girart}, \& {Rao}}]{Lai02}
---. 2002, \apj, 566, 925

\bibitem[{{Lazarian} \& {Hoang}(2007)}]{LazarianHoang07}
{Lazarian}, A., \& {Hoang}, T. 2007, \mnras, 378, 910

\bibitem[{{Matsakos} {et~al.}(2016){Matsakos}, {Tzeferacos}, \&
  {K{\"o}nigl}}]{Matsakos16}
{Matsakos}, T., {Tzeferacos}, P., \& {K{\"o}nigl}, A. 2016, \mnras, 463, 2716

\bibitem[{{Min} {et~al.}(2016){Min}, {Rab}, {Woitke}, {Dominik}, \&
  {M{\'e}nard}}]{Min16}
{Min}, M., {Rab}, C., {Woitke}, P., {Dominik}, C., \& {M{\'e}nard}, F. 2016,
  \aap, 585, A13

\bibitem[{{Miyake} \& {Nakagawa}(1993)}]{MiyakeNakagawa93}
{Miyake}, K., \& {Nakagawa}, Y. 1993, \icarus, 106, 20

\bibitem[{{Nagai} {et~al.}(2016){Nagai}, {Nakanishi}, {Paladino}, {Hull},
  {Cortes}, {Moellenbrock}, {Fomalont}, {Asada}, \& {Hada}}]{Nagai16}
{Nagai}, H., {Nakanishi}, K., {Paladino}, R., {et~al.} 2016, \apj, 824, 132

\bibitem[{{Okuzumi} {et~al.}(2016){Okuzumi}, {Momose}, {Sirono}, {Kobayashi},
  \& {Tanaka}}]{Okuzumi16}
{Okuzumi}, S., {Momose}, M., {Sirono}, S.-i., {Kobayashi}, H., \& {Tanaka}, H.
  2016, \apj, 821, 82

\bibitem[{{Pinte} {et~al.}(2016){Pinte}, {Dent}, {M{\'e}nard}, {Hales}, {Hill},
  {Cortes}, \& {de Gregorio-Monsalvo}}]{Pinte16}
{Pinte}, C., {Dent}, W.~R.~F., {M{\'e}nard}, F., {et~al.} 2016, \apj, 816, 25

\bibitem[{{Planck Collaboration} {et~al.}(2016){Planck Collaboration}, {Ade},
  {Aghanim}, {Alves}, {Arnaud}, {Arzoumanian}, {Aumont}, {Baccigalupi},
  {Banday}, {Barreiro}, {Bartolo}, {Battaner}, {Benabed}, {Benoit-L{\'e}vy},
  {Bernard}, {Bern{\'e}}, {Bersanelli}, {Bielewicz}, {Bonaldi}, {Bonavera},
  {Bond}, {Borrill}, {Bouchet}, {Boulanger}, {Bracco}, {Burigana}, {Calabrese},
  {Cardoso}, {Catalano}, {Chamballu}, {Chiang}, {Christensen}, {Clements},
  {Colombi}, {Colombo}, {Combet}, {Couchot}, {Crill}, {Curto}, {Cuttaia},
  {Danese}, {Davies}, {Davis}, {de Bernardis}, {de Rosa}, {de Zotti},
  {Delabrouille}, {Dickinson}, {Diego}, {Donzelli}, {Dor{\'e}}, {Douspis},
  {Ducout}, {Dupac}, {Elsner}, {En{\ss}lin}, {Eriksen}, {Falgarone},
  {Ferri{\`e}re}, {Finelli}, {Forni}, {Frailis}, {Fraisse}, {Franceschi},
  {Frejsel}, {Galeotta}, {Galli}, {Ganga}, {Ghosh}, {Giard},
  {Giraud-H{\'e}raud}, {Gjerl{\o}w}, {Gonz{\'a}lez-Nuevo}, {G{\'o}rski},
  {Gregorio}, {Gruppuso}, {Guillet}, {Hansen}, {Hanson}, {Harrison},
  {Hern{\'a}ndez-Monteagudo}, {Herranz}, {Hildebrandt}, {Hivon}, {Hobson},
  {Holmes}, {Huffenberger}, {Hurier}, {Jaffe}, {Jaffe}, {Jones}, {Juvela},
  {Keskitalo}, {Kisner}, {Knoche}, {Kunz}, {Kurki-Suonio}, {Lagache},
  {Lamarre}, {Lasenby}, {Lawrence}, {Leonardi}, {Levrier}, {Liguori}, {Lilje},
  {Linden-V{\o}rnle}, {L{\'o}pez-Caniego}, {Lubin}, {Mac{\'{\i}}as-P{\'e}rez},
  {Maffei}, {Mandolesi}, {Mangilli}, {Maris}, {Martin},
  {Mart{\'{\i}}nez-Gonz{\'a}lez}, {Masi}, {Matarrese}, {Mazzotta},
  {Melchiorri}, {Mendes}, {Mennella}, {Migliaccio}, {Mitra},
  {Miville-Desch{\^e}nes}, {Moneti}, {Montier}, {Morgante}, {Mortlock},
  {Munshi}, {Murphy}, {Naselsky}, {Nati}, {Natoli}, {N{\o}rgaard-Nielsen},
  {Noviello}, {Novikov}, {Novikov}, {Oppermann}, {Pagano}, {Pajot}, {Paladini},
  {Paoletti}, {Pasian}, {Perrotta}, {Pettorino}, {Piacentini}, {Piat},
  {Pierpaoli}, {Pietrobon}, {Plaszczynski}, {Pointecouteau}, {Polenta},
  {Pratt}, {Puget}, {Rachen}, {Rebolo}, {Reinecke}, {Remazeilles}, {Renault},
  {Renzi}, {Ricciardi}, {Ristorcelli}, {Rocha}, {Rosset}, {Rossetti},
  {Roudier}, {Rubi{\~n}o-Mart{\'{\i}}n}, {Rusholme}, {Sandri}, {Savelainen},
  {Savini}, {Scott}, {Soler}, {Stolyarov}, {Sutton}, {Suur-Uski}, {Sygnet},
  {Tauber}, {Terenzi}, {Toffolatti}, {Tomasi}, {Tristram}, {Tucci},
  {Valenziano}, {Valiviita}, {Van Tent}, {Vielva}, {Villa}, {Wade}, {Wandelt},
  {Yvon}, {Zacchei}, \& {Zonca}}]{PlanckXXXIII}
{Planck Collaboration}, {Ade}, P.~A.~R., {Aghanim}, N., {et~al.} 2016, \aap,
  586, A136

\bibitem[{{Pohl} {et~al.}(2016){Pohl}, {Kataoka}, {Pinilla}, {Dullemond},
  {Henning}, \& {Birnstiel}}]{Pohl16}
{Pohl}, A., {Kataoka}, A., {Pinilla}, P., {et~al.} 2016, \aap, 593, A12

\bibitem[{{Rao} {et~al.}(2014){Rao}, {Girart}, {Lai}, \& {Marrone}}]{Rao14}
{Rao}, R., {Girart}, J.~M., {Lai}, S.-P., \& {Marrone}, D.~P. 2014, \apjl, 780,
  L6

\bibitem[{{Rao} {et~al.}(2009){Rao}, {Girart}, {Marrone}, {Lai}, \&
  {Schnee}}]{Rao09}
{Rao}, R., {Girart}, J.~M., {Marrone}, D.~P., {Lai}, S.-P., \& {Schnee}, S.
  2009, \apj, 707, 921

\bibitem[{{Rebull} {et~al.}(2004){Rebull}, {Wolff}, \& {Strom}}]{Rebull04}
{Rebull}, L.~M., {Wolff}, S.~C., \& {Strom}, S.~E. 2004, \aj, 127, 1029

\bibitem[{{Ricci} {et~al.}(2010){Ricci}, {Testi}, {Natta}, {Neri}, {Cabrit}, \&
  {Herczeg}}]{Ricci10a}
{Ricci}, L., {Testi}, L., {Natta}, A., {et~al.} 2010, \aap, 512, A15

\bibitem[{{Segura-Cox} {et~al.}(2015){Segura-Cox}, {Looney}, {Stephens},
  {Fern{\'a}ndez-L{\'o}pez}, {Kwon}, {Tobin}, {Li}, \&
  {Crutcher}}]{Segura-Cox15}
{Segura-Cox}, D.~M., {Looney}, L.~W., {Stephens}, I.~W., {et~al.} 2015, \apjl,
  798, L2

\bibitem[{{Stephens} {et~al.}(2013){Stephens}, {Looney}, {Kwon}, {Hull},
  {Plambeck}, {Crutcher}, {Chapman}, {Novak}, {Davidson}, {Vaillancourt},
  {Shinnaga}, \& {Matthews}}]{Stephens13}
{Stephens}, I.~W., {Looney}, L.~W., {Kwon}, W., {et~al.} 2013, \apjl, 769, L15

\bibitem[{{Stephens} {et~al.}(2014){Stephens}, {Looney}, {Kwon},
  {Fern{\'a}ndez-L{\'o}pez}, {Hughes}, {Mundy}, {Crutcher}, {Li}, \&
  {Rao}}]{Stephens14}
---. 2014, Nature, 514, 597

\bibitem[{{Takahashi} \& {Inutsuka}(2014)}]{TakahashiInutsuka14}
{Takahashi}, S.~Z., \& {Inutsuka}, S.-i. 2014, \apj, 794, 55

\bibitem[{{Tazaki} {et~al.}(2017){Tazaki}, {Lazarian}, \& {Nomura}}]{Tazaki17}
{Tazaki}, R., {Lazarian}, A., \& {Nomura}, H. 2017, \apj, 839, 56

\bibitem[{{Tazaki} {et~al.}(2016){Tazaki}, {Tanaka}, {Okuzumi}, {Kataoka}, \&
  {Nomura}}]{Tazaki16}
{Tazaki}, R., {Tanaka}, H., {Okuzumi}, S., {Kataoka}, A., \& {Nomura}, H. 2016,
  \apj, 823, 70

\bibitem[{{Tomisaka}(2011)}]{Tomisaka11}
{Tomisaka}, K. 2011, \pasj, 63, 147

\bibitem[{{Ward-Thompson} {et~al.}(2017){Ward-Thompson}, {Pattle}, {Bastien},
  {Furuya}, {Kwon}, {Lai}, {Qiu}, {Berry}, {Choi}, {Coud{\'e}}, {Di Francesco},
  {Hoang}, {Franzmann}, {Friberg}, {Graves}, {Greaves}, {Houde}, {Johnstone},
  {Kirk}, {Koch}, {Kwon}, {Lee}, {Li}, {Matthews}, {Mottram}, {Parsons}, {Pon},
  {Rao}, {Rawlings}, {Shinnaga}, {Sadavoy}, {van Loo}, {Aso}, {Byun},
  {Chakali}, {Chen}, {Chen}, {Chen}, {Ching}, {Cho}, {Chrysostomou}, {Chung},
  {Doi}, {Drabek-Maunder}, {Eyres}, {Fiege}, {Friesen}, {Fuller}, {Gledhill},
  {Griffin}, {Gu}, {Hasegawa}, {Hatchell}, {Hayashi}, {Holland}, {Inoue},
  {Inutsuka}, {Iwasaki}, {Jeong}, {Kang}, {Kang}, {Kang}, {Kawabata}, {Kemper},
  {Kim}, {Kim}, {Kim}, {Kim}, {Kim}, {Kim}, {Lacaille}, {Lee}, {Lee}, {Li},
  {Li}, {Liu}, {Liu}, {Liu}, {Liu}, {Lyo}, {Mairs}, {Matsumura},
  {Moriarty-Schieven}, {Nakamura}, {Nakanishi}, {Ohashi}, {Onaka}, {Peretto},
  {Pyo}, {Qian}, {Retter}, {Richer}, {Rigby}, {Robitaille}, {Savini}, {Scaife},
  {Soam}, {Tamura}, {Tang}, {Tomisaka}, {Wang}, {Wang}, {Whitworth}, {Yen},
  {Yoo}, {Yuan}, {Zhang}, {Zhang}, {Zhou}, {Zhu}, {Andr{\'e}}, {Dowell},
  {Falle}, \& {Tsukamoto}}]{Ward-Thompson17}
{Ward-Thompson}, D., {Pattle}, K., {Bastien}, P., {et~al.} 2017, ArXiv
  e-prints, arXiv:1704.08552

\bibitem[{{Woitke} {et~al.}(2016){Woitke}, {Min}, {Pinte}, {Thi}, {Kamp},
  {Rab}, {Anthonioz}, {Antonellini}, {Baldovin-Saavedra}, {Carmona}, {Dominik},
  {Dionatos}, {Greaves}, {G{\"u}del}, {Ilee}, {Liebhart}, {M{\'e}nard},
  {Rigon}, {Waters}, {Aresu}, {Meijerink}, \& {Spaans}}]{Woitke16}
{Woitke}, P., {Min}, M., {Pinte}, C., {et~al.} 2016, \aap, 586, A103

\bibitem[{{Yang} {et~al.}(2016){Yang}, {Li}, {Looney}, \& {Stephens}}]{Yang16a}
{Yang}, H., {Li}, Z.-Y., {Looney}, L., \& {Stephens}, I. 2016, \mnras, 456,
  2794

\end{thebibliography}
\end{document}